\documentclass[journal,twoside,web]{ieeecolor}
\usepackage{tmi}
\usepackage{cite}
\usepackage{amsmath,amssymb,amsfonts}
\usepackage{algorithmic}
\usepackage{graphicx}
\usepackage{textcomp}
\usepackage{times}
\usepackage{cite}
\usepackage{amssymb}
\usepackage{amsmath,epsfig}
\usepackage{subcaption}
\usepackage{float}
\usepackage{siunitx}
\usepackage{array}
\usepackage{setspace}
\usepackage{indentfirst}
\usepackage{algorithm}
\usepackage{algorithmic}
\usepackage{bm}
\usepackage{caption}
\usepackage{graphicx,xcolor}
\usepackage{amsmath}
\usepackage{color}
\usepackage{stfloats}

\renewcommand{\vec}[1]{\mathbf{#1}}

\DeclareMathAlphabet{\mathpzc}{OT1}{pzc}{m}{it}
\newcommand{\vecs}[1]{\boldsymbol{#1}}
\makeatletter
\newcommand*{\rom}[1]{\expandafter\@slowromancap\romannumeral #1@}
\makeatother
\usepackage{float}
\usepackage{multicol}

\usepackage{hyperref}
\usepackage{adjustbox}
\usepackage{multirow}
\usepackage{multicol}

\def\BibTeX{{\rm B\kern-.05em{\sc i\kern-.025em b}\kern-.08em
    T\kern-.1667em\lower.7ex\hbox{E}\kern-.125emX}}
\begin{document}
\title{Ideal Observer Computation by Use of Markov-Chain Monte Carlo with Generative Adversarial Networks}

\author{Weimin Zhou,
	Umberto Villa,
        and Mark A. Anastasio
        \thanks{
This work was supported in part by NIH Awards
EB031772 (subproject 6366) and EB031585.
 (Corresponding author: Mark A. Anastasio and Weimin Zhou)}
\thanks{Weimin Zhou is with the Global Institute of Future Technology, Shanghai Jiao Tong University, 
Shanghai, China e-mail: weimin.zhou@sjtu.edu.cn.}
\thanks{Umberto Villa is with the Oden Institute for Computational Engineering \& Sciences,
The University of Texas at Austin, Austin, TX, 78712  USA e-mail: uvilla@oden.utexas.edu.}
\thanks{Mark A. Anastasio is with the Department 
of Bioengineering, University of Illinois Urbana-Champaign, Urbana,
IL, 61801 USA e-mail: maa@illinois.edu.}
}
\maketitle

\begin{abstract}
Medical imaging systems are often evaluated and optimized via objective, or task-specific, measures of image quality (IQ)
that quantify the performance of an observer on a specific clinically-relevant task.
The performance of the Bayesian Ideal Observer (IO)  sets an upper  limit among all observers, numerical or human, and has been advocated for use as
a figure-of-merit (FOM) for evaluating and optimizing medical imaging systems.
However, the IO test statistic corresponds to the likelihood ratio that is intractable to compute in the majority of cases.
A sampling-based method that employs Markov-Chain Monte Carlo (MCMC) techniques was previously proposed to estimate the IO performance.
However, current applications of MCMC methods for IO approximation have been limited to a  small number of situations
where the considered distribution of to-be-imaged objects can be described by a relatively simple stochastic object model (SOM).  
As such, there remains an important need to extend the domain of applicability of MCMC methods to address a large variety 
of scenarios where IO-based assessments are needed but the associated SOMs have not been available.
In this study, a novel MCMC method that employs a generative adversarial network (GAN)-based SOM, referred to as MCMC-GAN, is described and evaluated.
The MCMC-GAN method was quantitatively validated by use of test-cases for which reference solutions were available.
The results demonstrate that the MCMC-GAN method can extend the domain of applicability of MCMC methods for conducting IO analyses
of medical imaging systems.
\end{abstract}


\vspace{-.4cm}
\section{Introduction}
\label{sec:introduction}
It has been widely accepted that medical imaging systems should be evaluated and optimized based on objective measures of image quality (IQ) \cite{metz1995toward,barrett2013foundations,vennart1997icru}.
Objective measures of IQ quantify the ability of an observer to perform specific tasks that can be signal detection tasks  or parameter estimation tasks\cite{barrett2013foundations, kupinski2003ideal , shen2006using, zhou2020markov}.
A variety of observers have been actively explored as a means to compute objective measures of IQ for signal detection tasks\cite{barrett2015task,reiser2010task,sanchez2014task,glick2002investigation, barrett1998stabilized,gallas2003validating,park2007channelized, park2008markov, park2009efficient, zhou2019approximating, zhou2018learning, zhou2020approximate, zhou2019learningHO, he2020learning, granstedt2020learning, sidky2020signal, phillips2021hybrid, li2021supervised, li2021hybrid, li2021assessing}. 
Of these, the Bayesian Ideal Observer (IO) is distinct because it employs complete task-specific information and sets an upper performance limit among all observers \cite{myers2000ideal,barrett2013foundations,geisler2003ideal}. 
As such, the IO performance has been advocated for use as a figure-of-merit (FOM) when evaluating imaging systems \cite{vennart1997icru,metz1995toward}.
By use of this FOM, an imaging system can be optimized such that the amount of task-specific information present in the acquired image data is maximized. 
The IO performance can also be employed to compute the efficiency of other sub-optimal numerical observers or human observers~\cite{burgess1981efficiency}. 

When binary signal detection tasks are considered, the IO employs a likelihood ratio for computing the test statistic \cite{barrett2013foundations}.
The IO for a binary signal detection task provides the best possible 
receiver operating characteristic (ROC) curve for the given task and maximizes the
 area under the ROC curve (AUC)  \cite{barrett2013foundations, wagner1985unified}.
However, except in certain special cases, the IO test statistic cannot be described by a linear function of the image data and cannot be determined analytically.
Consequently, the literature on IO analyses of imaging systems has been largely limited to background-known-exactly (BKE) signal detection tasks and simple background-known-statistically (BKS) signal detection tasks that consider Gaussian backgrounds \cite{myers2000ideal,abbey2008ideal,jha2013ideal,anastasio2010analysis}.

Recently, the use of supervised learning for approximating the IO \cite{kupinski2001ideal} has been revisited and advanced with significant success to address more complicated detection and detection-localization tasks \cite{zhou2019learningIO,zhou2020approximate}.
While promising, this line of research remains ongoing and there remain tasks, such as certain detection-estimation tasks \cite{li2021hybrid}, which cannot be addressed by pure supervised learning methods and require the use of alternative statistical sampling methods \cite{li2021hybrid}.

In a seminal work by Kupinski \emph{et. al}, a sampling-based method that employs Markov-Chain Monte Carlo (MCMC) techniques was established to numerically approximate the IO test statistic when certain stochastic object models (SOMs), such as the lumpy object SOM \cite{rolland1992effect}, are considered \cite{kupinski2003ideal}. A SOM is a generative model that can be employed to sample from a prescribed statistical distribution of to-be-imaged objects \cite{kupinski2003experimental,zhou2022learning}.  A canonical use case
for a SOM is to produce an ensemble of objects, or discrete approximations of them, to enable computation of objective measures of image quality  via virtual imaging studies \cite{badano2021silico,samei2020virtual}.
This MCMC method was later adapted to estimate the IO test statistic for SOMs that include a binary texture model \cite{abbey2008ideal} and a parametrized torso phantom \cite{he2008toward}. The method was also employed 
to compute certain channelized IOs (CIOs) \cite{park2007channelized,park2008markov}.
However,  MCMC-based methods for use with more general SOMs have not been investigated and, moreover,
 the available SOMs  address only a  subset of medical imaging modalities and applications.
This currently limits the use of MCMC methods for performing IO analyses of
medical imaging systems.

Deep generative models, such as generative adversarial networks (GANs) \cite{goodfellow2014generative}, hold great potential for establishing SOMs that describe finite-dimensional approximations of objects.
A GAN comprises a generator and a discriminator that are both represented by deep neural networks.
The generator and   discriminator are trained jointly through an adversarial process.
Recent advances in GANs, such as progressively growing GANs (ProGANs) \cite{karras2017progressive} and style-based GANs \cite{karras2019style, karras2019analyzing}, have enabled the synthesis of high-resolution images.
When a GAN is specifically trained to sample from the distribution of   to-be-imaged objects,
the generator can be employed as a SOM \cite{zhou2019learning, zhou2021advancing, zhou2022learning}.

As a follow-up to a recent work by Zhou and Anastasio \cite{zhou2020markov}, 
this paper describes and evaluates 
a novel MCMC method that employs a GAN-based SOM, referred
to as MCMC-GAN. 
The MCMC-GAN method was designed to extend
the domain of applicability of MCMC methods for IO approximation to  applications in which the needed SOMs are not available but they could potentially be established by use of GANs or other deep generative models.
The MCMC-GAN method was quantitatively validated by use of test-cases for which reference solutions were available.
 Additionally, the method was applied to a problem for which no reference solution is available and traditional MCMC methods are not directly applicable.

The remainder of this work is organized as follows. In Sec. \ref{sec:bkgd},  the salient aspects of binary signal detection theory and traditional MCMC method for IO computation are reviewed. The proposed MCMC-GAN method is introduced in Sec. \ref{sec:method}. 
Numerical investigations and results of the MCMC-GAN method that involve
 a lumpy object model and a set of clinical MR brain images are provided in Secs. \ref{sec:ns} and \ref{sec:results}, respectively. Finally, the article concludes with a discussion of 
 potential advantages of the proposed MCMC-GAN method and topics for future study  in Sec. \ref{sec:con}.

\section{Background}
\label{sec:bkgd}
\vspace{-.1cm}
Consider a binary signal detection task 
that requires an observer to classify image data recorded by an imaging system
as
satisfying either a signal-absent hypothesis ($H_0$) or a signal-present hypothesis ($H_1$).
The measured image data under these hypotheses can be described as:
\begin{equation}\label{eq:cpt2_imgH}
\begin{split}
&H_{0}:  \mathbf{g} =  \mathbf{H}\mathbf{f}_b+ \mathbf{n}, \\
&H_{1}:  \mathbf{g} =  \mathbf{H}(\mathbf{f}_b + \mathbf{f}_s)+ \mathbf{n}, 
\end{split}
\end{equation}
where $\vec{g} \in \mathbb{R}^M$ denotes the measured image data acquired by a digitial imaging system, $\vec{n}\in \mathbb{R}^M$ denotes the measurement noise, and $\mathbf{H}$ denotes an imaging operator that maps objects to image data.
The quantities $\mathbf{f}_b$ and $\mathbf{f}_s$ denote the background object and to-be-detected signal, respectively.
For convenience, the imaged versions of the background object and signal will be denoted as $\vec{b} \equiv  \mathbf{H}\mathbf{f}_b$ and $\vec{s} \equiv  \mathbf{H}\mathbf{f}_s$, respectively.

The imaging operator $\mathbf{H}$ can describe different   mappings, depending on whether a continuous-to-discrete (C-D) or discrete-to-discrete (D-D) formulation of the forward problem is considered \cite{barrett2013foundations}. 
In the case where the imaging process is  described as a C-D mapping, which is the proper description of a digital imaging system, $\mathbf{f}_b$ and $\mathbf{f}_s$ are functions and $\mathbf{H}$ maps a function to the $M$-dimensional vector $\mathbf{g}$ that describes the measured image.
If the imaging process is approximated as a D-D mapping,
discretized approximations of $\mathbf{f}_b$ and $\mathbf{f}_s$ are described as $N$-dimensional vectors and the matrix operator
$\mathbf{H} \in \mathbb{R}^{M\times N}$ maps the finite-dimensional representation of the object to  $\mathbf{g}$.

For a signal-known-exactly/background-known-statistically 
(SKE/BKS) detection task, the signal
$\mathbf{f}_s$ is deterministic but the background object $\mathbf{f}_b$ 
is random, being described by a probability density function (PDF) $\mathrm{pr}(\mathbf{f}_b)$ that characterizes the variability in the cohort of to-be-imaged subjects. Although one typically does not have direct access to this PDF, in principle, a SOM can be established to sample from it\cite{kupinski2003experimental,zhou2022learning}.
Accordingly, object variability and measurement noise are
the contributors to variability in the image data
$\mathbf{g}$ for a SKE/BKS detection task.
Such  tasks will be considered in this work.

To perform a binary signal detection task, an observer computes a test statistic $t(\vec{g})$ that maps the measured image data $\vec{g}$ to a real-valued scalar. 
The test statistic $t(\vec{g})$ is compared to a pre-determined threshold $\tau$ to classify $\vec{g}$ as belonging to one of the two hypotheses.
A ROC curve \cite{metz1986roc} that depicts the trade-off between the false-positive fraction (FPF)
and the true-positive fraction (TPF) can be plotted by varying the threshold $\tau$. The area under the ROC curve (AUC) can be subsequently computed to quantify the observer performance.

\subsection{Bayesian Ideal Observer}
\label{sec:BIO}
The Bayesian Ideal Observer (IO) implements a decision
strategy that employs complete statistical knowledge and sets an upper performance limit among all observers.
The test statistic employed by the IO for a binary signal detection task is any monotonic transformation of the likelihood ratio $\Lambda(\vec{g})$ defined as
\cite{barrett2013foundations, kupinski2003ideal, kupinski2001ideal}
\begin{equation}
\Lambda(\vec{g}) = \frac{\mathrm{pr}(\vec{g}|H_1)}{\mathrm{pr}(\vec{g}|H_0)},
\end{equation}
where $\mathrm{pr}(\vec{g}|H_j)$ is the conditional PDF that describes the likelihood of the degraded $\vec{g}$ under the hypothesis $H_j$ ($j=0, 1$).
Computation of the IO test statistic for BKS tasks is analytically intractable in the majority of cases because evaluation of the conditional PDFs 
$\mathrm{pr}(\vec{g}|H_j)$ involve $\mathrm{pr}(\mathbf{f}_b)$ that is not explicitly known.
In the case where a SOM is available to sample from $\mathrm{pr}(\mathbf{f}_b)$,
a Markov chain Monte Carlo (MCMC) method proposed by Kupinski \emph{et al.}  \cite{kupinski2003ideal} can be employed to approximate the IO test statistic as described next.

\vspace{-.2cm}
\subsection{Markov Chain Monte Carlo}
\label{sec:MCMC}
When a SKE binary signal detection task is considered, the likelihood ratio $\Lambda(\vec{g})$ can be computed as  \cite{kupinski2003ideal}:
\vspace{-.1cm}
\begin{multline} \label{eq:trad_mcmc}
\Lambda (\vec{g})  = \frac{\int d\vec{b}\ \mathrm{pr}_{b}(\vec{b}) \mathrm{pr}(\vec{g}|\vec{b}, H_1)}{\int d\vec{b}\ \mathrm{pr}_{b}(\vec{b}) \mathrm{pr}(\vec{g}|\vec{b}, H_0)} \equiv  \\
\int d\vec{b}\ \Lambda_{\text{BKE}} (\vec{g}|\vec{b}) \mathrm{pr}(\vec{b}|\vec{g}, H_0),
\end{multline}
where $\Lambda_{\text{BKE}} (\vec{g}|\vec{b})$ is the likelihood ratio given a background image data $\vec{b}=\mathbf{H}\mathbf{f}_b$ and $\mathrm{pr}(\vec{b}|\vec{g}, H_0)$ is a posterior probability density function. These quantities can be computed as:
\begin{subequations}
\label{eq:trad_mcmc2}
\begin{equation}
\Lambda_{\text{BKE}} (\vec{g}|\vec{b}) = \frac{\mathrm{pr}(\vec{g}|\vec{b},H_1)}{\mathrm{pr}(\vec{g}|\vec{b},H_0)},
\end{equation}
and
\begin{equation}
\mathrm{pr}(\vec{b}|\vec{g}, H_0) = \frac{\mathrm{pr}(\vec{g}|\vec{b}, H_0) \mathrm{pr}_b(\vec{b})}{\int d\vec{b'} \mathrm{pr}(\vec{g}|\vec{b'}, H_0) \mathrm{pr}_b(\vec{b'})},
\end{equation}
\end{subequations}
where $\mathrm{pr}_b(\vec{b})$ is the PDF of background image data $\vec{b}$.
Consider that the random background object $\vec{f}_b$ can be described by a SOM that is characterized by a random vector $\vecs{\theta}$ that has a PDF $\mathrm{pr}(\vecs{\theta})$.
In this case, the background image data $\vec{b}$ can be generated by sampling $\vecs{\theta}$: i.e., $\vec{b} \equiv \vec{b}(\vecs{\theta})$.
In terms of these quantities, the likelihood ratio described in Eq. (\ref{eq:trad_mcmc}) can be subsequently computed as  \cite{kupinski2003ideal}:
\begin{equation} 
\label{eq:trad_mcmc3}
\Lambda (\vec{g})  = \int d\vecs{\theta}\ \Lambda_{\text{BKE}} (\vec{g}|\vec{b}(\vecs{\theta})) \mathrm{pr}(\vecs{\theta}|\vec{g}, H_0).
\end{equation}
Monte Carlo integration can be  employed to approximate this integral to yield an estimate of the
 likelihood ratio \cite{kupinski2003ideal}:
\begin{equation}
{\Lambda}(\vec{g}) \approx \frac{1}{J}\sum_{j=1}^{J} \Lambda_{\text{BKE}} (\vec{g}|\vec{b}(\vecs{\theta}^{j})),
\end{equation}
where the samples $\vecs{\theta}^{j}$
are drawn from the posterior probability function  $
\mathrm{pr}(\vecs{\theta}|\vec{g}, H_0)$ and $J$ is the number of samples employed to approximate the integral.
To obtain the samples $\vecs{\theta}^{j}$,
a Markov chain having the stationary density $\mathrm{pr}(\vecs{\theta}|\vec{g},H_0)$ can be generated by use of a Metropolis-Hastings algorithm \cite{chib1995understanding}.
Specifically, an initial vector $\vecs{\theta}^0$ is selected and a proposal density function $q(\vecs{\tilde{\theta}} | \vecs{\theta}^j)$
is specified. For a given vector  $\vecs{\theta}^j$, $j\ge 1$, 
a candidate $\vecs{\tilde{\theta}}$ for the next sample in the chain  is sampled from the proposal density  $q(\vecs{\tilde{\theta}} | \vecs{\theta}^j)$
and is accepted with a probability $\mathrm{Pr}_\text{a}(\vecs{\tilde{\theta}} | \vecs{\theta}^j, \vec{g})$ that is defined as:
\begin{multline}
\mathrm{Pr}_\text{a}(\vecs{\tilde{\theta}}| \vecs{\theta}^j, \vec{g}) = \\
\min \left[1, \frac{\mathrm{pr}(\vec{g} | \vec{b}(\vecs{\tilde{\theta}}), H_0) \mathrm{pr}(\vecs{\tilde{\theta}}) q(\vecs{\theta}^j|\vecs{\tilde{\theta}}) }{\mathrm{pr}(\vec{g} | \vec{b}({\vecs{\theta}}^j), H_0) \mathrm{pr}({\vecs{\theta}}^j) q(\vecs{\tilde{\theta}} | \vecs{\theta}^j)}\right].
\end{multline}
If the candidate vector $\vecs{\tilde{\theta}}$ is accepted, it is added to the Markov chain: $\vecs{\theta}^{j+1} = \vecs{\tilde{\theta}}$;
otherwise, $\vecs{\theta}^{j+1} = \vecs{\theta}^j$.

However, current applications of MCMC methods have been limited to relatively simple SOMs such as a lumpy object model \cite{kupinski2003ideal}, a binary texture model \cite{abbey2008ideal}, and a parameterized torso phantom \cite{he2008toward}.  
Next, an extension of this method for use with GAN-based SOMs is
presented.

\subsection{Generative Adversarial Networks}
\label{sec:GAN}
Generative adversarial networks (GANs) have been actively explored and successfully applied to establish deep generative models
to generate new (``fake'') images that are consistent with the stochastic properties of ensembles of training (``real'') images \cite{goodfellow2014generative}. A GAN comprises a generator and a discriminator that are both represented by deep neural networks. 
The generator is trained against a discriminator through an adversarial process.
After the training, the generator can be employed to generate ``fake'' images that can represent statistical properties of ``real'' images.

When a GAN is trained on a set of finite-dimensional background objects $\vec{f}_b$,
the generator maps a random latent vector $\vec{z} \in \mathbb{R}^k$ to a ``fake'' background object $\vec{\hat{f}}_b = G(\vec{z}; \mathbf{\Theta}_G)$.
Here, $G(\cdot\ ; \mathbf{\Theta}_G): \mathbb{R}^k \rightarrow  \mathbb{R}^N$ is a mapping function represented by a deep neural network with a weight vector $\mathbf{\Theta}_G$, and the latent vector $\vec{z}$ is sampled from a simple known distribution such as normal distribution. 
The discriminator is represented by another deep neural network with a weight vector $\mathbf{\Theta}_D$ and a mapping function $D(\cdot\ ; \mathbf{\Theta}_D): \mathbb{R}^M \rightarrow  \mathbb{R}$. The discriminator maps an image to a real-valued score for use to distinguish between ``real" and ``fake" images. 
A GAN is trained by playing a two-player minimax game between the generator and the discriminator:
\begin{equation} \label{eq:GAN}
\begin{split}
\min_{\mathbf{\Theta}_G} \max_{\mathbf{\Theta}_D} V(D,G) = & {E_{ \vec{f}_b }} [l\left(D(\vec{b}; \mathbf{\Theta}_D)\right)]\\
 + &{E_{\vec{z} }} [l(1- D\left( G(\vec{z}; \mathbf{\Theta}_G); \mathbf{\Theta}_D \right) )],
\end{split}
\end{equation}
 where $l(\cdot)$ is an objective function, which is dependent on specific training strategies.
When $D(\cdot\ ; \mathbf{\Theta}_D)$ and  $G(\cdot\ ; \mathbf{\Theta}_G)$ possess sufficient capacity,  $P_{\vec{\hat{f}}_b} = P_{\vec{f}_b}$ when the global optimum of the minimax game is achieved \cite{goodfellow2014generative}.
Here, $P_{\vec{f}_b}$ denotes the distribution of the ``real" background object $\vec{f}_b$, and $P_{\vec{\hat{f}}_b} $ denotes the distribution of the ``fake" background object $\vec{\hat{f}}_b$.
The generator can subsequently represent a SOM that describes the variability within the ensemble of background objects, and the background image data $\hat{\vec{b}}$ that is parameterized by the latent vector $\vec{z}$ can be computed as:
\begin{equation}
\label{eq:bhat1}
\hat{\vec{b}} = \mathbf{H}G(\vec{z}; \Theta_G) \equiv \hat{\vec{b}}(\vec{z}).
\end{equation}

A GAN sometimes may also be trained directly on background image data $\vec{b}$. In such cases, the generator can directly generate ``fake'' background image data:
\begin{equation}
\label{eq:bhat2}
\vec{\hat{b}}: \vec{\hat{b}} = G(\vec{z}; \Theta_G) \equiv \vec{\hat{b}}(\vec{z}).
\end{equation}
This is particularly useful when a CD imaging operator is considered such that the GAN cannot be directly applied to establish the object model.

\section{Markov-Chain Monte Carlo approximation of the IO by use of GANs}
\label{sec:method}
Here, the MCMC method described in Sec. \ref{sec:MCMC} is generalized for use with the GAN-generated  data $\vec{\hat{b}}$ defined in Eq. \ref{eq:bhat1} or \ref{eq:bhat2}, depending on which object model is used.
Consider a SKE/BKS signal detection task, similar to the Eqs. \ref{eq:trad_mcmc}---\ref{eq:trad_mcmc3}, the IO test statistic for a degraded image data $\vec{\hat{g}}$ can be computed as:
\begin{equation}
\Lambda (\vec{\hat{g}}) 
= \int d\vec{z} \Lambda_{\text{BKE}} (\vec{\hat{g}}|\vec{\hat{b}}(\vec{z})) p(\vec{z}|\vec{\hat{g}}, H_0),
\end{equation}
where $\Lambda_{\text{BKE}} (\vec{\hat{g}}|\vec{\hat{b}}(\vec{z})) $ and $p(\vec{z}|\vec{\hat{g}}, H_0)$ can be computed as:
\begin{subequations}
\begin{equation}
\Lambda_{\text{BKE}} (\vec{\hat{g}}|\vec{\hat{b}}(\vec{z})) = \frac{p(\vec{\hat{g}}|\vec{\hat{b}}(\vec{z}),H_1)}{p(\vec{\hat{g}}|\vec{\hat{b}}(\vec{z}),H_0)}.
\end{equation}
\begin{equation}
p(\vec{z}|\vec{\hat{g}}, H_0)= \frac{p(\vec{\hat{g}}|\vec{\hat{b}}(\vec{z}),H_0) p_z(\vec{z})}{\int d\vec{z'} p(\vec{\hat{g}}|\vec{\hat{b}}(\vec{z'}), H_0) p_z(\vec{z'})}.
\end{equation}
\end{subequations}
Markov chain Monte Carlo simulation can be subsequently employed to approximate the likelihood ratio:
\begin{equation}
\label{eq:mcmcgan}
{\Lambda}(\vec{\hat{g}}) \approx \frac{1}{J}\sum_{j=1}^{J} \Lambda_{\text{BKE}} (\vec{\hat{g}}|\vec{\hat{b}}(\vec{z}^j)),
\end{equation}
where $\vec{z}^j$ is sampled from the posterior distribution $p(\vec{z}|\vec{\hat{g}}, H_0)$. 
To construct a Markov chain that draws samples from the posterior distribution $p(\vec{z}|\vec{\hat{g}}, H_0)$,
a proposal density function needs to be specified. 
Given the current sample $\vec{z}^j$, a candidate latent vector $\vec{\tilde{z}}$ is proposed by sampling from the proposal density function $q(\vec{\tilde{z}}|\vec{z}^j)$
and is accepted to the Markov chain with the acceptance probability: 
\begin{equation}\label{eq:ap}
p_\text{a}(\vec{\tilde{z}} | \vec{z}^j, \vec{\hat{g}}) = \min \left[1, \frac{p\big(\vec{\hat{g}} | \vec{\hat{b}}(\vec{\tilde{z}}), H_0\big) p_z(\vec{\tilde{z}}) q(\vec{z}^j|\vec{\tilde{z}}) }{p\big(\vec{\hat{g}} | \vec{\hat{b}}({\vec{z}}^j), H_0\big) p_z({\vec{z}}^j) q(\vec{\tilde{z}} | \vec{z}^j)}\right].
\end{equation}
Here, the probability density function $p_{z}(\cdot)$ has a simple analytical form because the latent vector $\vec{z}$ is sampled from a known distribution such
as the normal distribution.

In this study, the latent vector $\vec{z}$ is sampled from a normal distribution: $\vec{z}\sim \mathcal{N}(0, \vec{I}_k)$. The proposal density function $q(\vec{\tilde{z}}|\vec{z}^j)$ was designed based on the preconditioned Crank–Nicolson (pCN) algorithm \cite{cotter2013mcmc}.
Specifically, given the current latent vector  $\vec{z}^j$, the candidate vector $\vec{\tilde{z}}$ is computed as:
\begin{equation}
\label{eq:pCN1}
\vec{\tilde{z}} =  \sqrt{1-\beta ^2}\vec{z}^j + \beta \vecs{\xi}, 
\end{equation} 
where $\beta$ is the step size and $\vecs{\xi}$ is sampled from normal distribution $\mathcal{N}(0, \vec{I}_k)$.
Given the prior invariance of the pCN proposal in Eq. \eqref{eq:pCN1}, i.e.
\begin{equation}
\label{eq:pCN2}
p_z(\vec{\tilde{z}}) q(\vec{z}^j|\vec{\tilde{z}}) = p_z({\vec{z}}^j) q(\vec{\tilde{z}} | \vec{z}^j),
\end{equation} 
the acceptance probability in Eq. \ref{eq:ap} can be computed as
\begin{equation}\label{eq:ap2}
p_\text{a}(\vec{\tilde{z}} | \vec{z}^j, \vec{\hat{g}}) = \min \left[1, \frac{p\big(\vec{\hat{g}} | \vec{\hat{b}}(\vec{\tilde{z}}), H_0\big) }{p\big(\vec{\hat{g}} | \vec{\hat{b}}({\vec{z}}^j), H_0\big) }\right].
\end{equation}

 \section{Numerical studies}
 \label{sec:ns}
 Computer-simulation studies were conducted to investigate the ability of the proposed MCMC-GAN method to approximate the IO test statistic associated with SOMs that are established by use of GANs.
 Two SKE/BKS binary signal detection tasks corresponding to SOMs of different levels of realism were considered.
In the first numerical study, the proposed MCMC-GAN method was applied to objects produced by a lumpy object model. The IO performance was validated by use of the conventional MCMC algorithm that was designed specifically for lumpy object models \cite{kupinski2003ideal}.
 The second numerical study applies the proposed MCMC-GAN method to a set of clinical brain MR images that cannot be described by an existing SOM to which the conventional MCMC method can be readily applied.
  The observer performance was assessed by use of the ROC curve. The Metz-ROC software \cite{metz1998rockit} was used for curve fitting with the ``proper'' binormal model~\cite{metz1999proper, pesce2007reliable}.
 Details of the two considered signal detection tasks are provided below.
 
   \subsection{Signal detection task with lumpy background}
   The first SKE/BKS binary signal detection task employed a stochastic lumpy object model to simulate the random background. The so-called lumpy background (LB) can be described as:
   \begin{equation}
	f_b(\vec{r}) = \sum_{n=1}^{N_b}l(\vec{r}-\vec{r}_n|a, w_b),
\end{equation} 
where $N_b$ is the random number of lumps that follows a Poisson distribution with the mean of 6,  and $l(\vec{r}-\vec{r}_n|a, w_b)$ describes the shape of lumps that was modeled by a 2D Gaussian function:
\begin{equation}
l(\vec{r}-\vec{r}_n|a, w_b) = {a}\exp\left(-\frac{(\vec{r}-\vec{r}_n)^T(\vec{r}-\vec{r}_n)}{2w_b^2}  \right).
\end{equation}
Here, $a = 1$, $w_b = 8$, and $\vec{r}_n$ denotes the center location of the $n^{th}$ lump. The lump location was sampled from a uniform distribution over the image field of view {of $64\times 64$}.

The signal was modeled by a 2D Gaussian function:
\begin{equation}
\label{eq:signal}
f_{s}(\vec{r}) = {a_{s}}\exp\left(-\frac{(\vec{r}-\vec{r}_s)^T(\vec{r}-\vec{r}_s)}{2w_s^2}  \right),
\end{equation}
where $a_s = 0.3$ is the signal amplitude, $w_s = 2.5$ is the signal width, and $\vec{r}_s = [32, 32]^T$ is the signal location corresponding to the center of the field of view. 

The lumpy model characterizes continuous objects.
An idealized parallel-hole collimator imaging system that can be described by a linear C-D mapping was considered 
to produce degraded images.
This imaging system can be described as a convolution with a Gaussian point response function (PRF)~\cite{kupinski2003ideal, kupinski2003experimental}:
\begin{equation}
h_m(\vec{r}) = \frac{\mathpzc{h}}{2\pi w_h^2}\exp\left(-\frac{(\vec{r}-\vec{r}_m)^T(\vec{r}-\vec{r}_m)}{2w_h^2}  \right),
\end{equation}
where $h_m(\vec{r}) $ is the PRF that describes the sensitivity of the $m^{th}$ element in the degraded  image to the object at the location $\vec{r}$, $\mathpzc{h} = 35$ and $w_h = 2$ are the height and width of the PRF, respectively. The virtual imaging system acquires images of the size $64\times 64$. The $m^{th}$ $(1\leq m \leq 4096)$ element of the background image data ${b}_m$ and that of the signal image data $s_m$ can be calculated as:
\begin{equation}
b_{m} = \frac{a\mathpzc{h}w_b^2}{w_h^2+w_b^2} \sum_{n=1}^{N_b }\exp\left(-\frac{(\vec{r}_m-\vec{r}_n)^T(\vec{r}_m-\vec{r}_n)}{2(w_h^2+w_s^2)}  \right), \\
\end{equation}
and
\begin{equation}
s_m =  \frac{a_s\mathpzc{h}w_b^2}{w_h^2+w_s^2}\exp\left(-\frac{(\vec{r}_m-\vec{r}_s)^T(\vec{r}_m-\vec{r}_s)}{2(w_h^2+w_b^2)}  \right).
\end{equation}

\begin{figure}[ht!]
  \centering
 \includegraphics[width=1.0\linewidth]{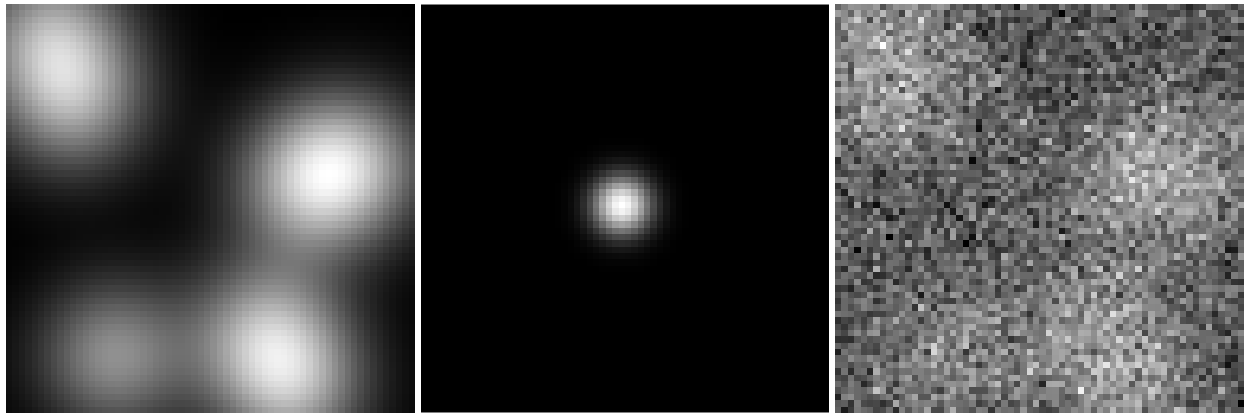}
 \caption{Left: A lumpy background image data. Middle: The signal image data corresponding to the considered signal detection task. Right: A signal-present noisy measured image.}
 \label{fig:LB_sig}
\end{figure}

The noise was modeled by independent and identically distributed (i.i.d.) multivariate Gaussian distribution with the mean of 0 and standard deviation of 20. 
Examples of the noiseless background image data $\vec{b}$, the signal image data $\vec{s}$, and the signal-present measurement data $\vec{g}$ are shown in Fig. \ref{fig:LB_sig}.

A GAN that employs a progressive growing training strategy (i.e., ProGAN) \cite{karras2017progressive} was trained by use of 10,000 samples of background image data. After the training, the generator with the input vector $\vec{z}$ can be employed to synthesize background image data and 
the IO test statistic can be computed according to Eq. \ref{eq:mcmcgan}.
Because i.i.d Gaussian noise was considered, the BKE likelihood ratio can be computed as:
\begin{equation}
\label{eq:lb_lambda}
\begin{split}
\Lambda_{\text{BKE}} (\hat{\vec{g}}| \hat{\vec{b}}(\vec{z}^j) ) 
=  \exp{\left[(\hat{\vec{g}} - \hat{\vec{b}}(\vec{z}^j) - \vec{s}/2)^T K^{-1}_n\vec{s} \right]},
\end{split}
\end{equation}
 where $K_n$ is the covariance matrix corresponding to i.i.d. Gaussian noise with standard deviation of 20.
To compute the IO test statistic,
 the Markov chain was constructed by use of pCN algorithm with an acceptance probability $p_\text{a}(\vec{\hat{z}} | \vec{z}^j, \vec{g}) $ that is the ratio of Gaussian likelihood functions:
 \begin{equation}\label{eq:eq_acpt}
 \begin{split}
&p_\text{a}(\vec{\tilde{z}} | \vec{z}^j, \vec{g}) = \\
\min &\left(1, \frac{ \exp{\left[-\frac{1}{2}(\hat{\vec{g}} - \hat{\vec{b}}(\vec{\tilde{z}}))^TK^{-1}_n(\hat{\vec{g}} - \hat{\vec{b}}(\vec{\tilde{z}})) \right]}}{ \exp{\left[-\frac{1}{2}(\hat{\vec{g}} - \hat{\vec{b}}(\vec{z}^j))^TK^{-1}_n(\hat{\vec{g}} - \hat{\vec{b}}(\vec{z}^j)) \right]}}\right).
\end{split}
\end{equation}

 The proposed MCMC-GAN method was applied to 200 signal-absent and 200 signal-present images to estimate the IO performance.  
 The conventional MCMC method that was developed for lumpy backgrounds (MCMC-LB) \cite{kupinski2003ideal} was implemented to provide the reference performance of the IO for validation.
 As a further validation, the Hotelling observer (HO) performance was computed for the considered lumpy background by use of a covariance matrix decomposition  method \cite{barrett2013foundations}.

  \subsection{Signal detection task with clinical MRI images}
\begin{figure}[th!]
  \centering
 \includegraphics[width=1.0\linewidth]{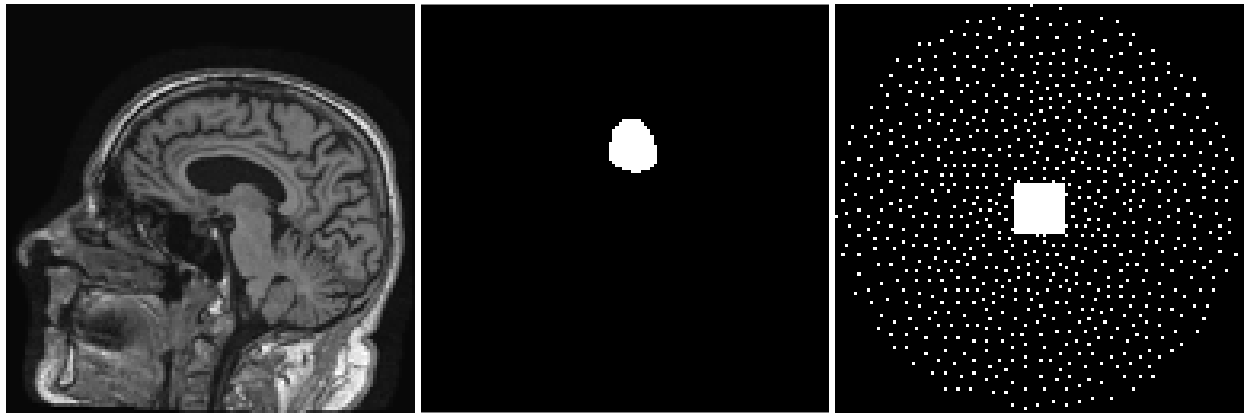}
 \caption{Left: An example of the MR brain object $\vec{f}_b$. Middle: The considered tumor signal to be detected. Right: The k-space sampling pattern.}
 \label{fig:MR_sig}
\end{figure}

  In this study, a SKE/BKS binary signal detection task that considers clinical brain MR images was considered.
  A clinical brain MR dataset sponsored by Alzheimer's Disease Neuroimaging Initiative (ADNI) \cite{mueller2005alzheimer}
 was employed as the set of ground-truth background objects $\vec{f}_b$. Twelve thousand high quality sagittal brain MR images were selected and resized to the dimension of $128\times 128$.
 These images were subsequently normalized between 0 and 1 for use as training images for training a ProGAN.  
 After the training, the generator of the ProGAN was employed to synthesize the background object by sampling the random latent vector $\vec{z}$.
A signal object with realistic brain tumor shape was considered that was selected from a brain tumor dataset (\url{https://figshare.com/articles/dataset/brain_tumor_dataset/1512427}).
A stylized MR imaging system that acquires undersampled k-space data was considered and the variable-density Poisson-disc sampling pattern \cite{lustig2007sparse, uecker2015berkeley} with an acceleration factor of 16 was employed. {A finite-dimensional approximation of objects was considered in this case and the imaging operator was described by a D-D mapping.}
The measurement data $\vec{g}$ were simulated by adding the measurement noise to the undersampled k-space data that were computed by use of a 2D discrete Fourier transform (DFT).
The measurement noise $\vec{n}$ was modeled by i.i.d. zero mean complex Gaussian random vector with a standard deviation $\sigma$ of 80 for both the real and imaginary components. {The peak signal-to-noise ratio (PSNR), which is defined as $PSNR=20\times \log_{10}(\frac{MAX_g}{\sigma})$, was 34.89 dB. Here, $MAX_g$ denotes the maximum value of the measurement data $\vec{g}$ evaluated on the testing dataset.}
An example of the considered MR brain images $\vec{f}_b$, the to-be-detected signal $\vec{f}_s$, and the k-space sampling pattern are shown in Fig. \ref{fig:MR_sig}.
 
The proposed MCMC-GAN method was applied to the k-space data, and the IO performance was 
evaluated on 200 signal-absent and 
200 signal-present measurement data.
The pCN algorithm was employed to construct a Markov chain for each image.
Because i.i.d. Gaussian noise was considered, the BKE likelihood ratio and the proposal acceptance probability were computed according to Eqs. \ref{eq:lb_lambda} and \ref{eq:eq_acpt}, respectively.

Because the considered MR images cannot be described by an existing SOM to which the MCMC method can be applied, the ground-truth IO performance was not provided.
However,
the supervised learning method that employs convolutional neural networks (CNNs) \cite{zhou2019approximating} was implemented to provide a reference IO performance.
When training CNNs, a training dataset that comprised one million samples of ProGAN-generated background image data and a “semi-online learning” method \cite{zhou2019approximating}  in which the measurement data were generated on-the-fly were employed.
A multi-channel CNN having 15  convolutional (CONV) layers was applied to the complex-valued inverse DFT of the zero-filled k-space data to estimate the IO performance.
Each CONV layer comprised 32 filters with 5 × 5 spatial support and was followed by a LeakyReLU activation function. The last CONV layer was followed by a max-pooling layer and a fully connected (FC) layer.
An extra validation was provided by computing the HO performance by use of a covariance matrix decomposition method.

  \subsection{Convergence analysis of MCMC-GAN}
  The potential scale reduction factor (PSFR) is a widely used metric to monitor MCMC convergence for a scalar variable of interest \cite{gelman1992inference, brooks1998general, Vehtari2020}.
The PSFR is calculated on parallel chains and measures the ratio of the averaged variance of within-sequence samples to the variance of the pooled samples across parallel chains. 
Let $M>1$ denote the number of chains and $N_c$ denote the number of samples in each chain,
the PSFR can be computed as \cite{gelman1992inference, brooks1998general, Vehtari2020}:
\begin{equation}
\text{PSFR} = \sqrt{\frac{N_c-1}{N_c} + \frac{1}{N_c}\frac{B}{W}},
\end{equation}
where $W$ is the within-sequence variance and $B$  is the between-sequence variance.
These quantities  are given by:
\begin{subequations}
\begin{equation}
W = \frac{1}{M(N_c -1)} \sum_{m=1}^{M}\sum_{n=1}^{N_c} (v^{n, m} - \bar{v}^{.,m})^2,
\end{equation}
\begin{equation}
\mathbf{B} = \frac{N_c}{M-1} \sum_{m=1}^{M} (\bar{v}^{.,m} -\bar{v}^{.,.})^2,
\end{equation}
\end{subequations}
where $v^{n,m}$ denotes the $n^{th}$ sample of the $m^{th}$ chain, $\bar{v}^{.,m}$ denotes the average of samples from the $m^{th}$ chain and  $\bar{v}^{.,.}$ denotes the average of all samples:
\begin{align}
\bar{v}^{.,m} & = \frac{1}{N_c} \sum_{n=1}^{N_c} v^{n, m}, & \bar{v}^{.,.}& = \frac{1}{M} \sum_{m=1}^{M} \bar{v}^{.,m}. 
\end{align}
The PSFR is a real-valued scalar that is always greater than or equal to one. 
When the PSFR approaches one, the Markov chain converges to a target distribution.
A threshold of 1.01 has been advocated for use as a threshold to determine the convergence of the Markov chains \cite{Vehtari2020}.

In this study, the PSFR was evaluated on the BKE likelihood ratio $\Lambda_\text{BKE}$, which is a scalar that was employed to compute the Monte Carlo integration for approximating the IO test statistic.
The PSFR was calculated by use of 
five parallel chains that were generated by running the MCMC-GAN with different random seeds.
 
 \subsection{MCMC-GAN implementation details}
The ProGANs were trained on 4 NVIDIA Quadro RTX 8000 GPUs by use of Tensorflow \cite{abadi2016tensorflow}.
A stochastic gradient method that employs the Adam algorithm \cite{kingma2014adam} was employed as the optimizer to train the ProGANs.
A ProGAN architecture with the initial image resolution 
of $4\times 4$  and a 64-dimensional latent space were employed. More details of the ProGAN architecture used in this study can be found in \cite{karras2017progressive}.
The ProGANs were trained by use of the publicly available ProGAN code (\url{https://github.com/tkarras/progressive\_growing\_of\_gans}).

 {After the ProGAN was trained, the MCMC technique with pCN proposal was employed to sample the GAN's latent variable for use in computing the IO test statistic. The corresponding proposal density function and the acceptance probability were defined in  Eq. \ref{eq:pCN1} and Eq. \ref{eq:pCN2}, respectively.} 
 Each Markov chain was run for 200,000 iterations on a single NVIDIA Quadro RTX 8000 GPU. A burn-in period corresponding to the first 10,000 iterations was discarded from each Markov chain. 
 The IO test statistic (i.e., likelihood ratio) corresponding to each degraded image data was subsequently computed by evaluating the Monte Carlo integration on the 190,000 iterations of the Markov chain according to Eq. \ref{eq:mcmcgan}. 
 For each considered signal detection task, the MCMC GAN was applied to a set of 200 signal-absent images and 200 signal-present images, and the resulting test statistics were employed to assess the IO performance.
 \section{Results}
\label{sec:results}
  \subsection{Signal detection task with lumpy background}
\begin{figure}[ht!]
  \centering
 \includegraphics[width=1.0\linewidth]{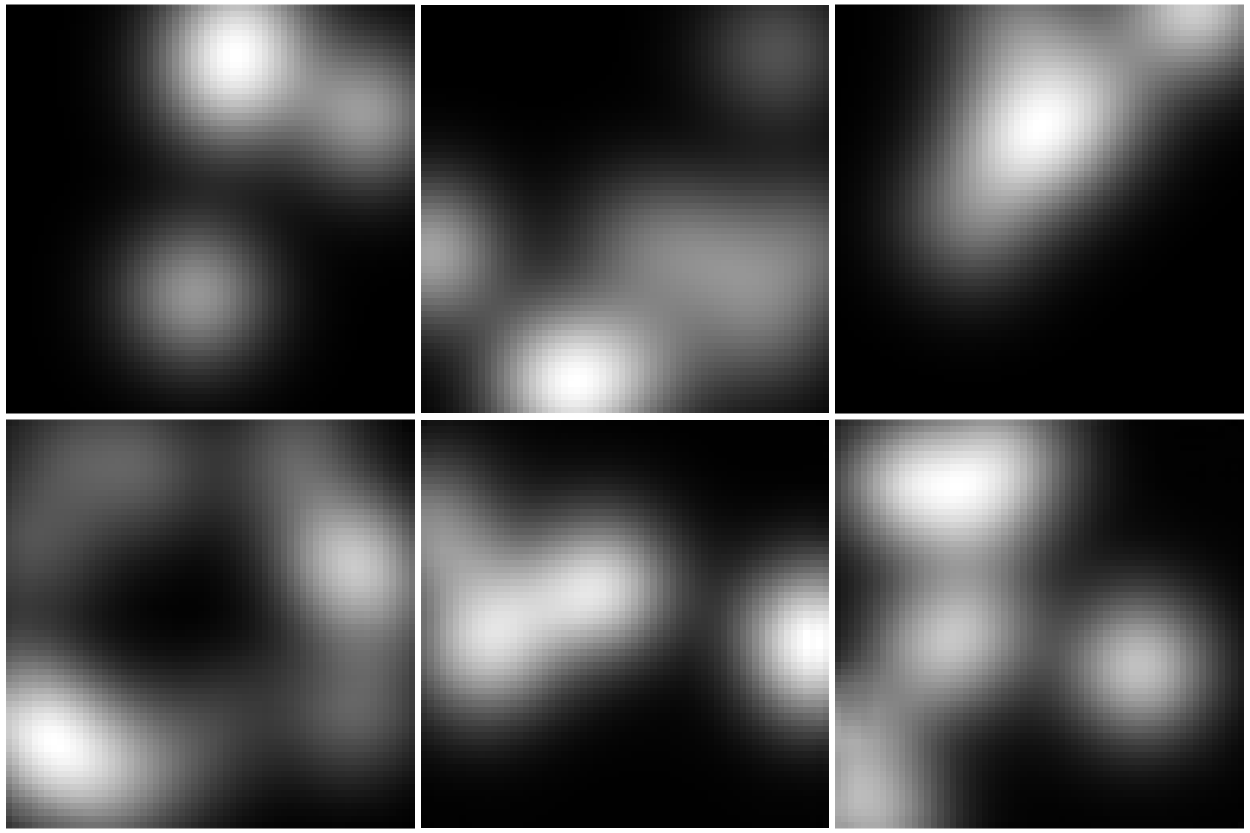}
 \caption{Top: Three examples of ground-truth lumpy background images. Bottom: Thee examples of ProGAN-generated images.}
 \label{fig:LB_imgs}
\end{figure}

  The ground-truth (top row) lumpy background images
  and the ProGAN-generated (bottom row) images are shown in Fig. \ref{fig:LB_imgs}.
 The ProGAN-generated images and the ground-truth lumpy background images have similar visual appearances.

\begin{figure}[ht!]
  \centering
 \includegraphics[width=1.0\linewidth]{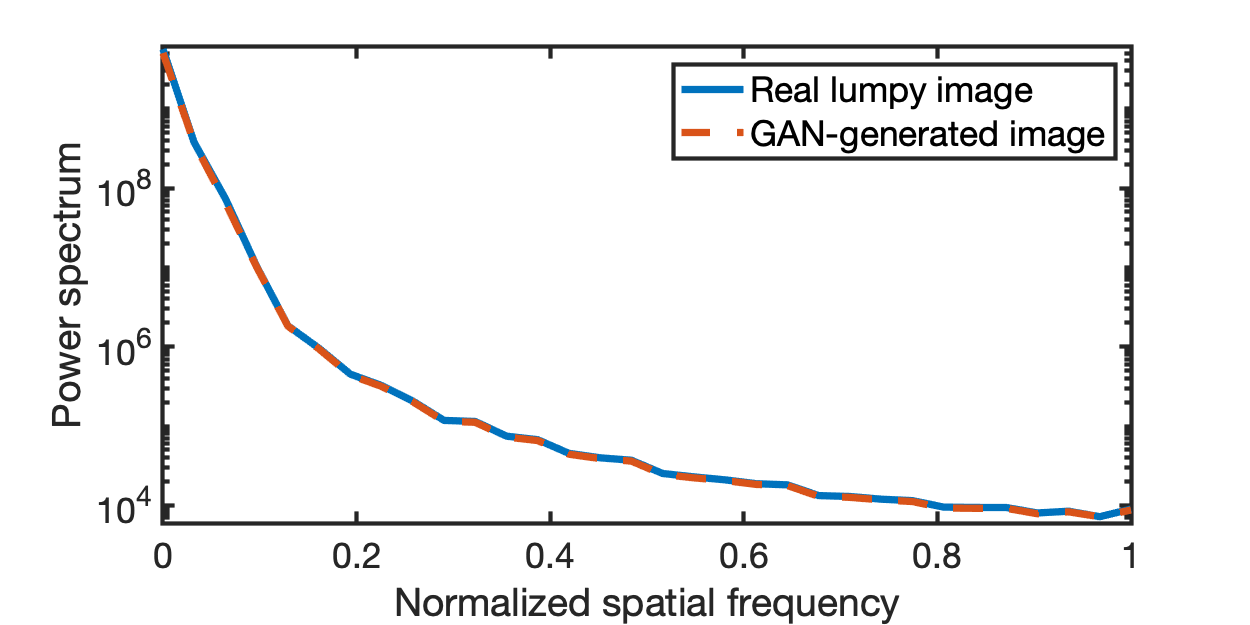}
 \caption{The power spectra of ``real" and ProGAN-generated images. The power spectrum of the ProGAN-generated images is almost identical to that of the ``real" images.}
 \label{fig:LB_PS}
\end{figure}

The power spectra {\cite{bochud1999statistical}} that describe the frequency content of the ``real" and ProGAN-generated ``fake" images were employed to assess the ProGAN.
The power spectrum of the ProGAN-generated images (red-dashed curve) is compared to that of the ``real" images (blue curve)  in Fig. \ref{fig:LB_PS}.
These power spectra were radially averaged over all angles, and averaged over 200 ``real" and 200 ``fake" lumpy background images, respectively. 
The two power spectra are almost identical.

\begin{figure}[ht!]
\centering
\begin{subfigure}[b]{0.45\textwidth}
   \centering
 \includegraphics[width=1.0\linewidth]{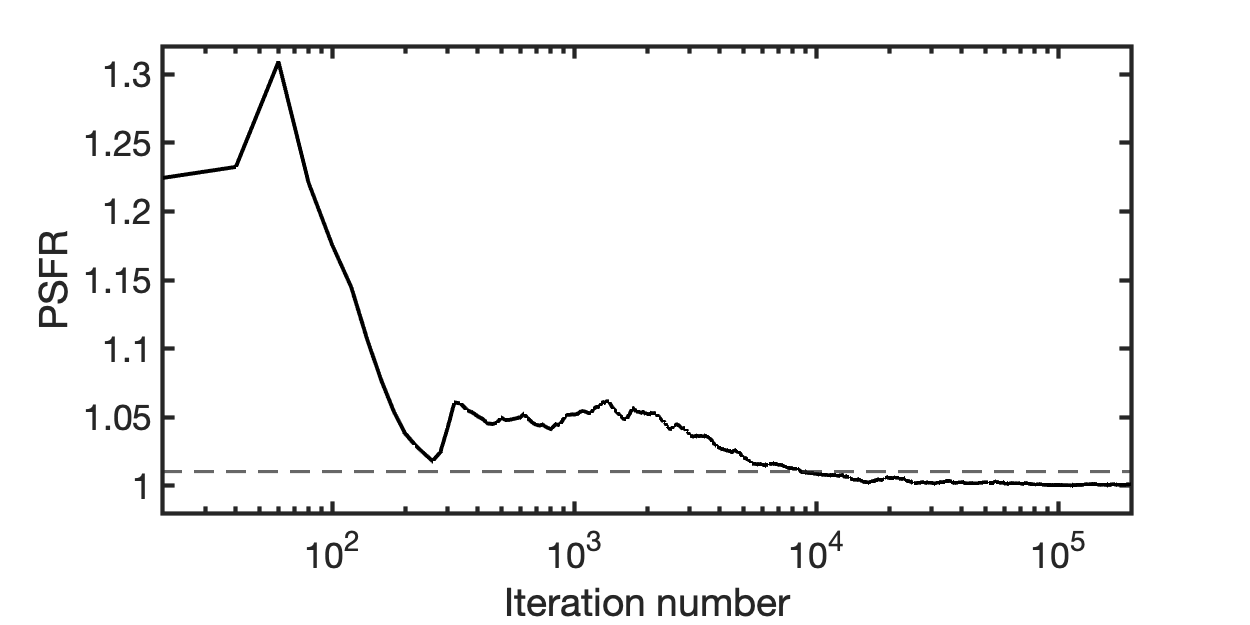}
 \caption{}
 \end{subfigure}
 \begin{subfigure}[b]{0.45\textwidth}
   \centering
 \includegraphics[width=1.0\linewidth]{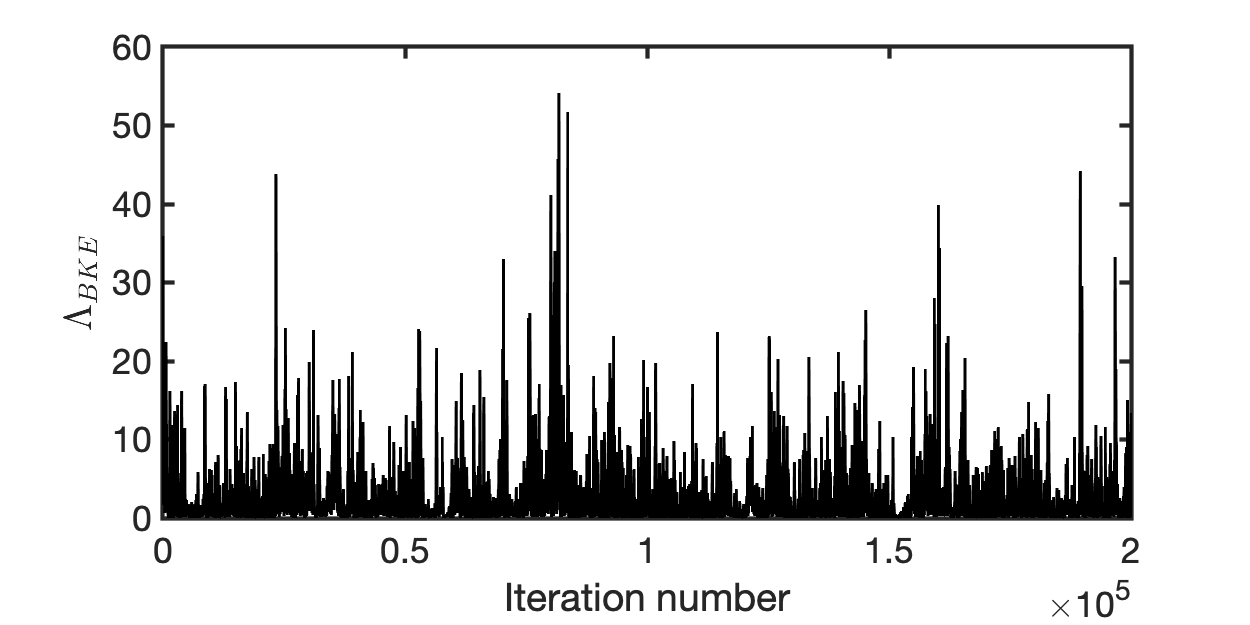}
 \caption{}
 \end{subfigure}
 \begin{subfigure}[b]{0.45\textwidth}
  \centering
 \includegraphics[width=1.0\linewidth]{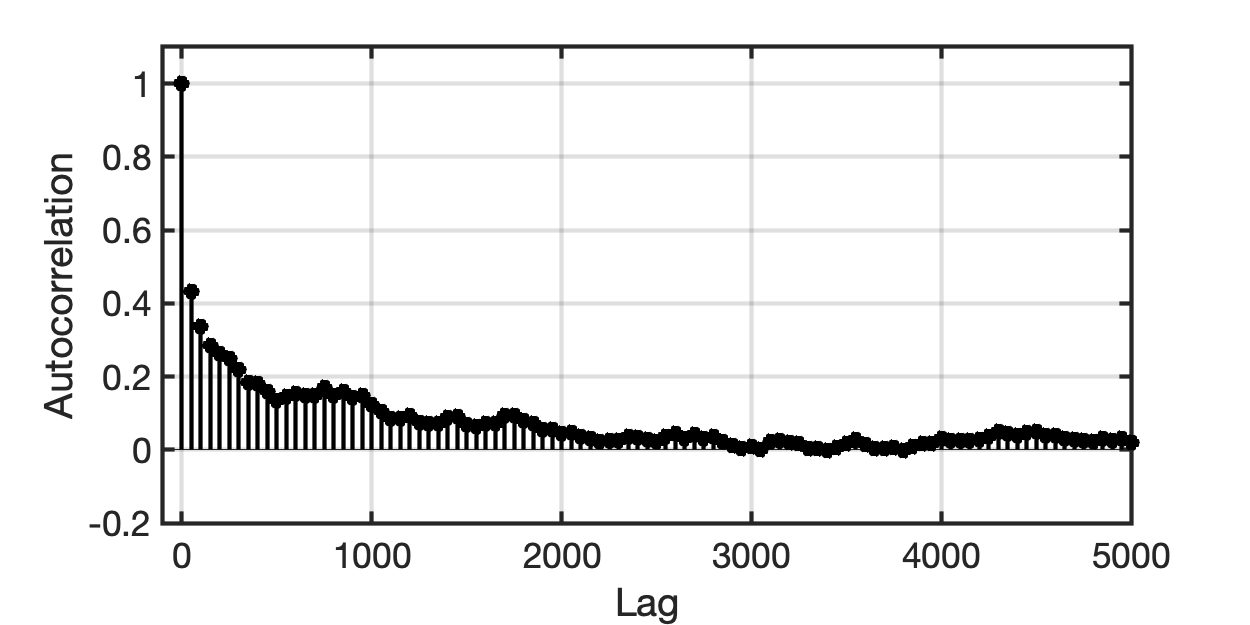}
 \caption{}
 \end{subfigure}
 \caption{(a) The curve of PSFR of $\Lambda_\text{BKE}$ as a function of iteration number of the Markov chain. The chain of $\Lambda_\text{BKE}$ converged after about 10,000 iterations. (b) A chain of $\Lambda_{BKE}$ evaluated at different iterations of the Markov chain. (c) The autocorrelation of the chain in (b).}
 \label{fig:diag_LB}
\end{figure}

The plot of the PSFR of the $\Lambda_\text{BKE}$ samples as a function of the iteration number of the Markov chain for a signal-present lumpy image is shown in Fig. \ref{fig:diag_LB} (a). The horizontal dashed line in Fig. \ref{fig:diag_LB} (a) indicates the convergence threshold of 1.01. 
The PSFR approached one when the iteration number increased, and converged after about 10,000 iterations.
 The final PSFR value at the end of the chain was 1.0008.
A chain of the BKE likelihood ratio $\Lambda_\text{BKE}$ evaluated at different iterations of the Markov chain is shown in Fig. \ref{fig:diag_LB} (b), and its autocorrelation function is plotted in Fig. \ref{fig:diag_LB} (c). 

\begin{figure}[ht!]
  \centering
 \includegraphics[width=1.0\linewidth]{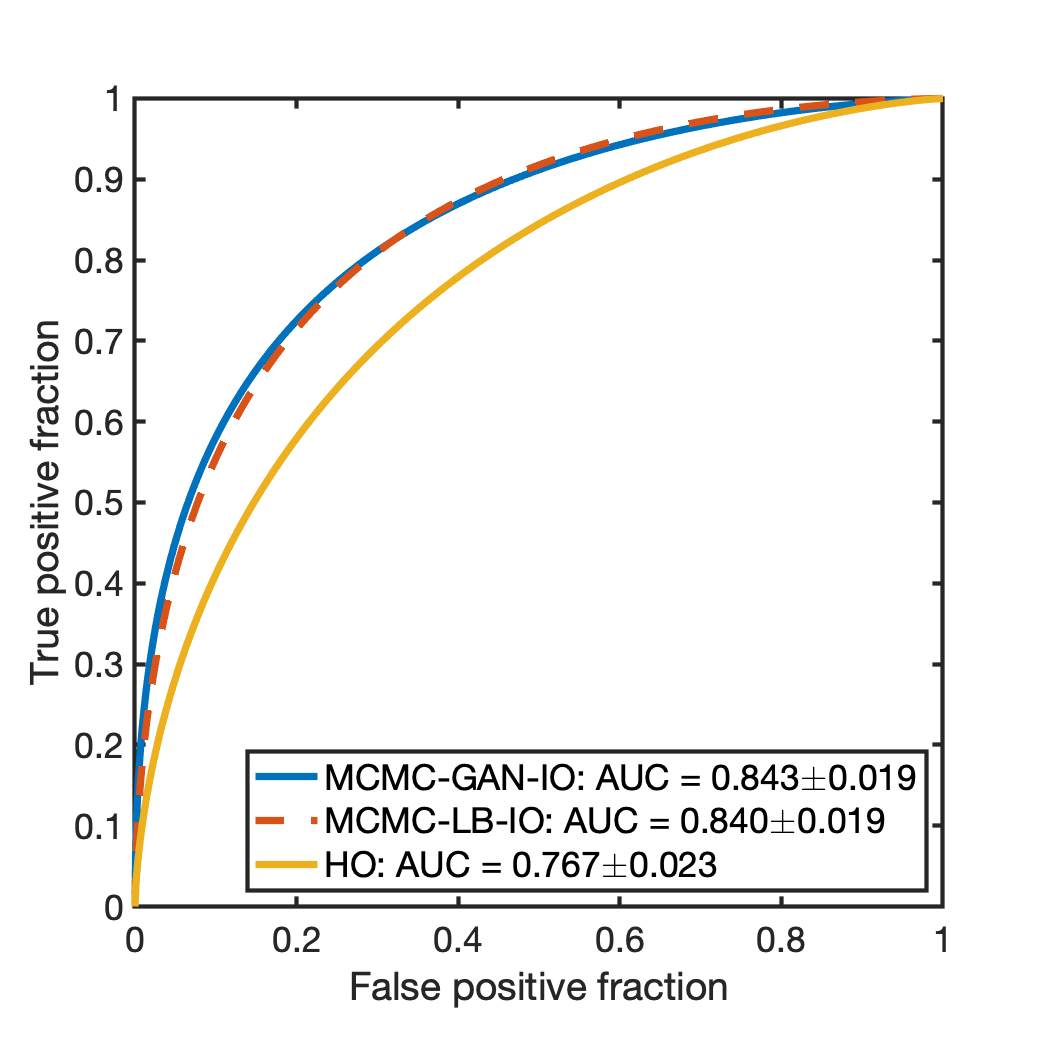}
 \caption{The ROC curves produced by the MCMC-GAN-IO, MCMC-LB-IO and the HO. The ROC curve corresponding to the MCMC-GAN-IO is in close agreement with the MCMC-LB-IO and is higher than the HO.}
 \label{fig:LB_roc}
\end{figure}

The ROC curves corresponding to the MCMC-GAN IO (blue curve), MCMC-LB IO (red-dashed curve) and the HO (yellow curve) are shown in Fig. \ref{fig:LB_roc}.
The AUC value corresponding to the MCMC-GAN IO, MCMC-LB IO and the HO are $0.843\pm {0.019}$, $0.840\pm {0.019}$ and $0.767\pm {0.023}$, respectively.
The performance of the MCMC-GAN IO is in close agreement with that of the MCMC-LB IO and is higher than that of the HO.

  \subsection{Signal detection task with clinical MR images}
\begin{figure}[ht!]
  \centering
 \includegraphics[width=1.0\linewidth]{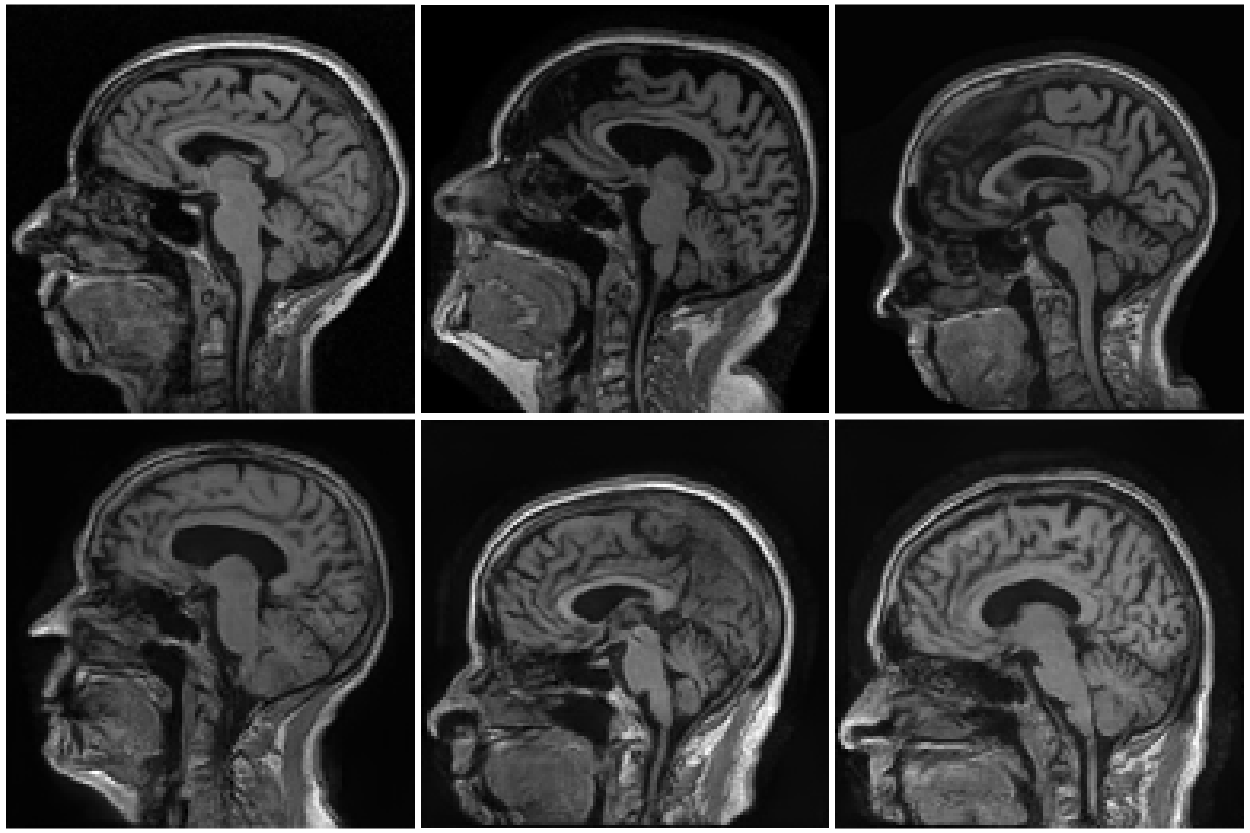}
 \caption{Top: Three  examples of ground-truth MR images. Bottom: Three examples of ProGAN-generated images.}
 \label{fig:MR_imgs}
\end{figure}

  The ground-truth (top row) clinical MR brain images
  and the ProGAN-generated (bottom row) images are shown in Fig. \ref{fig:MR_imgs}.
 The ProGAN-generated images and the ground-truth MR images have similar visual appearances.

\begin{figure}[ht!]
  \centering
 \includegraphics[width=1.0\linewidth]{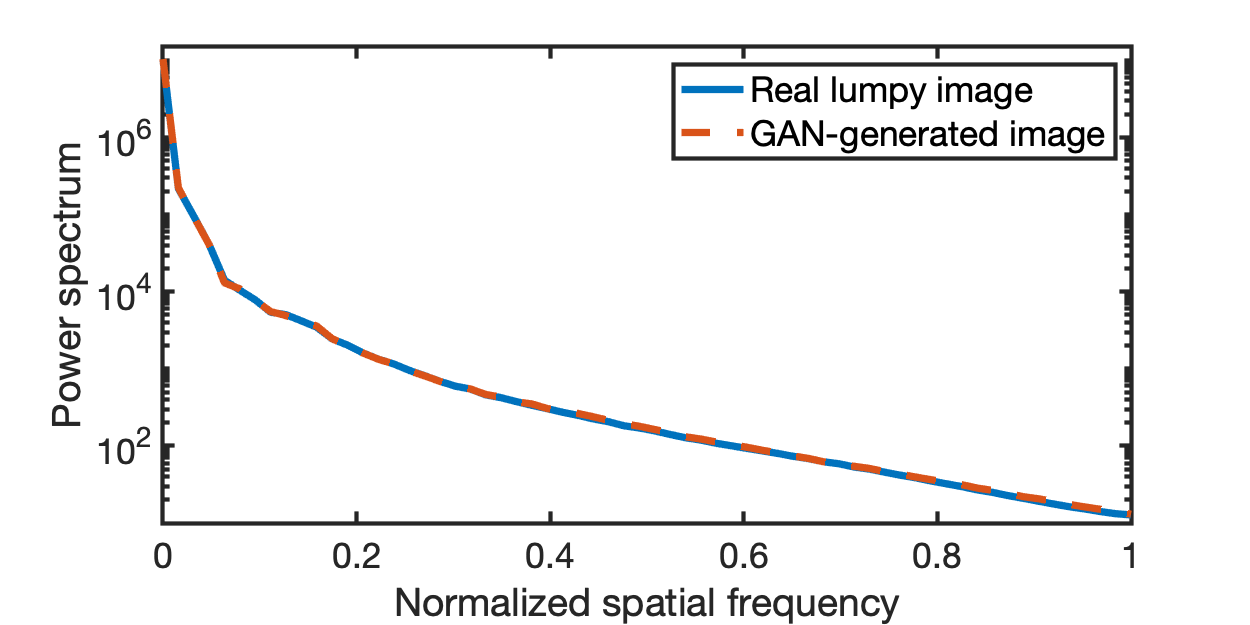}
 \caption{Power spectrums of ``real" and ``fake" images. These power spectrums are almost identical.}
 \label{fig:MR_PS}
\end{figure}

The power spectrum analysis is provided to evaluate the ProGAN.
The power spectra of ``real" clinical MR images (blue curve) and GAN-generated ``fake" MR images (red-dashed curve) are shown in Fig. \ref{fig:MR_PS}.
These power spectra were radially averaged over all angles, and averaged over 200 ``real" and 200 ``fake" MR brain images, respectively. 
The two power spectra are almost identical.

   \begin{figure}[h!]
\centering
\begin{subfigure}[b]{0.5\textwidth}
   \centering
 \includegraphics[width=1.0\linewidth]{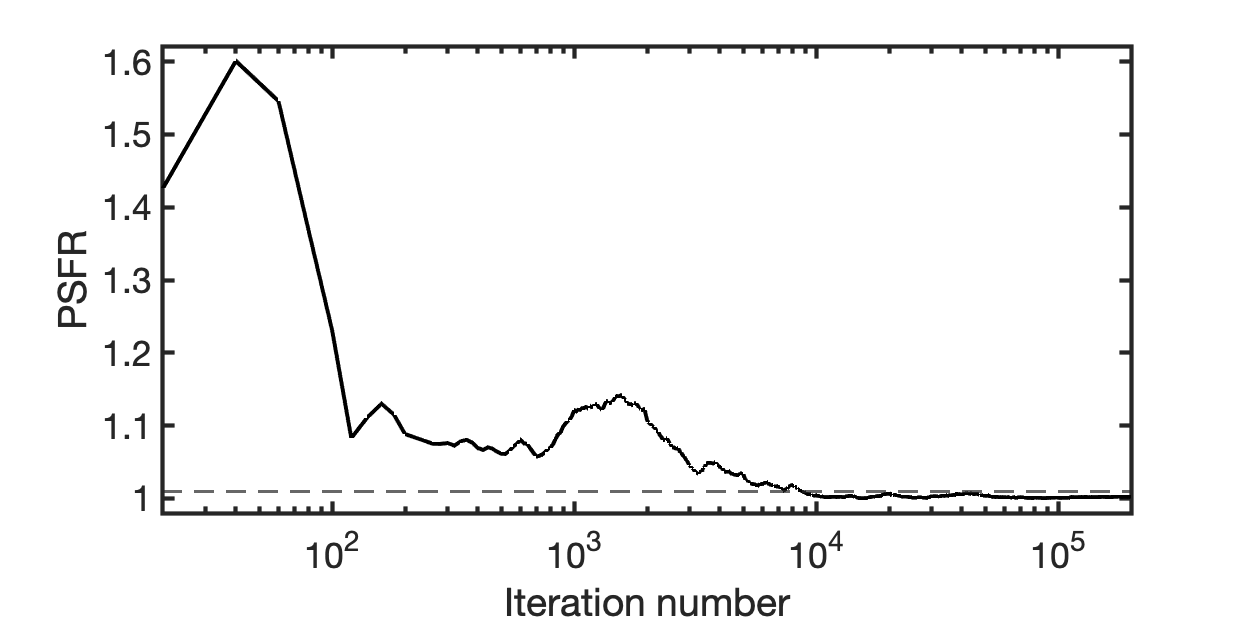}
 \caption{}
 \end{subfigure}
 \begin{subfigure}[b]{0.5\textwidth}
   \centering
 \includegraphics[width=1.0\linewidth]{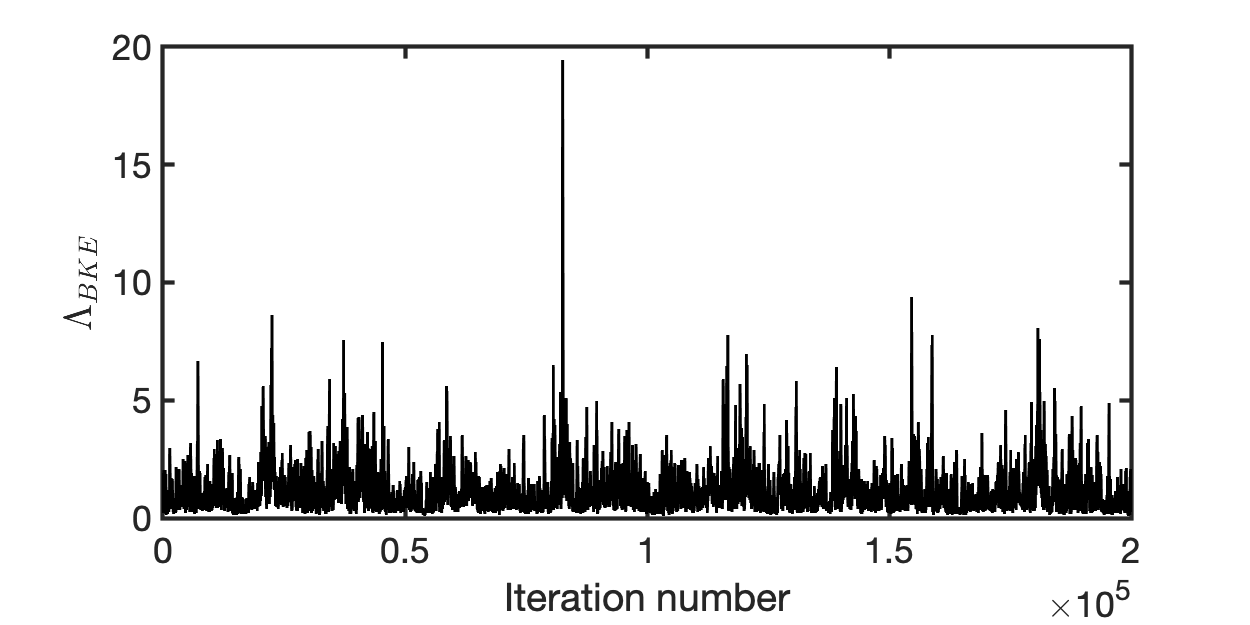}
 \caption{}
 \end{subfigure}
 \begin{subfigure}[b]{0.5\textwidth}
  \centering
 \includegraphics[width=1.0\linewidth]{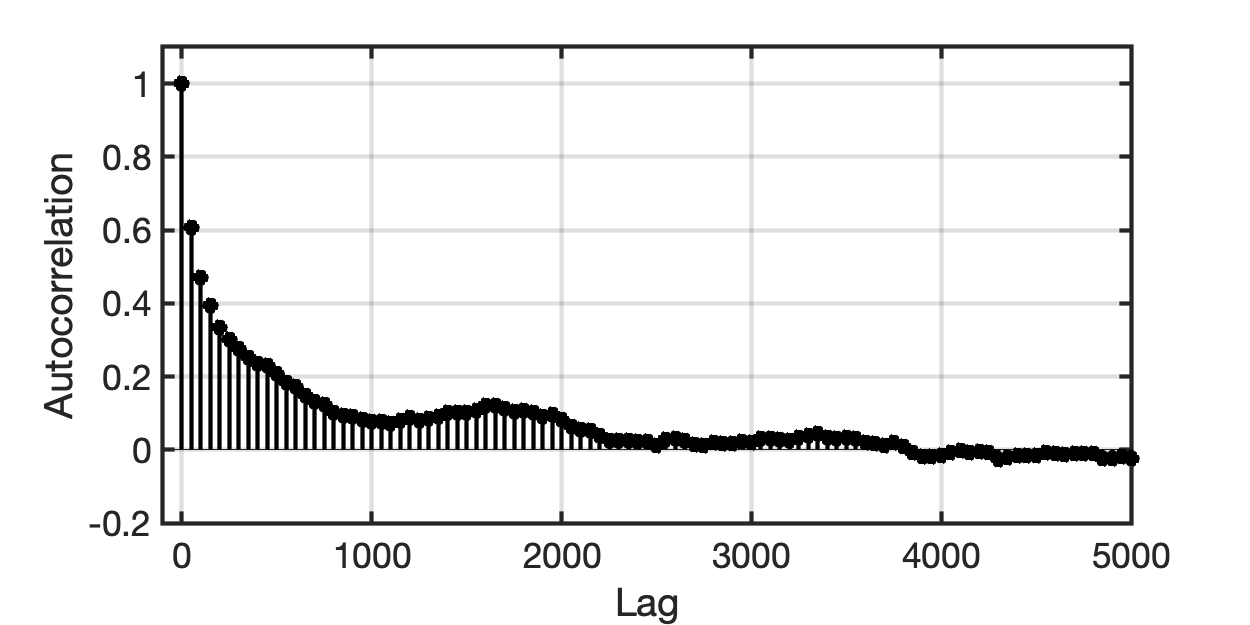}
 \caption{}
 \end{subfigure}
 \caption{(a) The curve of PSFR of $\Lambda_\text{BKE}$ as a function of iteration number of the Markov chain. The chain of $\Lambda_\text{BKE}$ converged after about 10,000 iterations. (b) A chain of $\Lambda_{BKE}$ evaluated at different iterations of the Markov chain. (c) The autocorrelation of the chain in (b).}
 \label{fig:diag_MR}
\end{figure}

The plot of the PSFR of $\Lambda_\text{BKE}$ as a function of the iteration number of the Markov chain for a signal-present MR brain image is shown in Fig. \ref{fig:diag_MR} (a). The horizontal dashed line in Fig. \ref{fig:diag_MR} (a) indicates the convergence threshold of 1.01. 
The PSFR approached one when the iteration number increased, and converged after about 10,000 iterations.
 The final PSFR value at the end of the chain was 1.0026.
A chain of the BKE likelihood ratio $\Lambda_\text{BKE}$ evaluated at different iterations of the Markov chain is shown in Fig. \ref{fig:diag_MR} (b).
The autocorrelation of the chain is plotted in Fig. \ref{fig:diag_MR} (c).

\begin{figure}[ht!]
  \centering
 \includegraphics[width=1.0\linewidth]{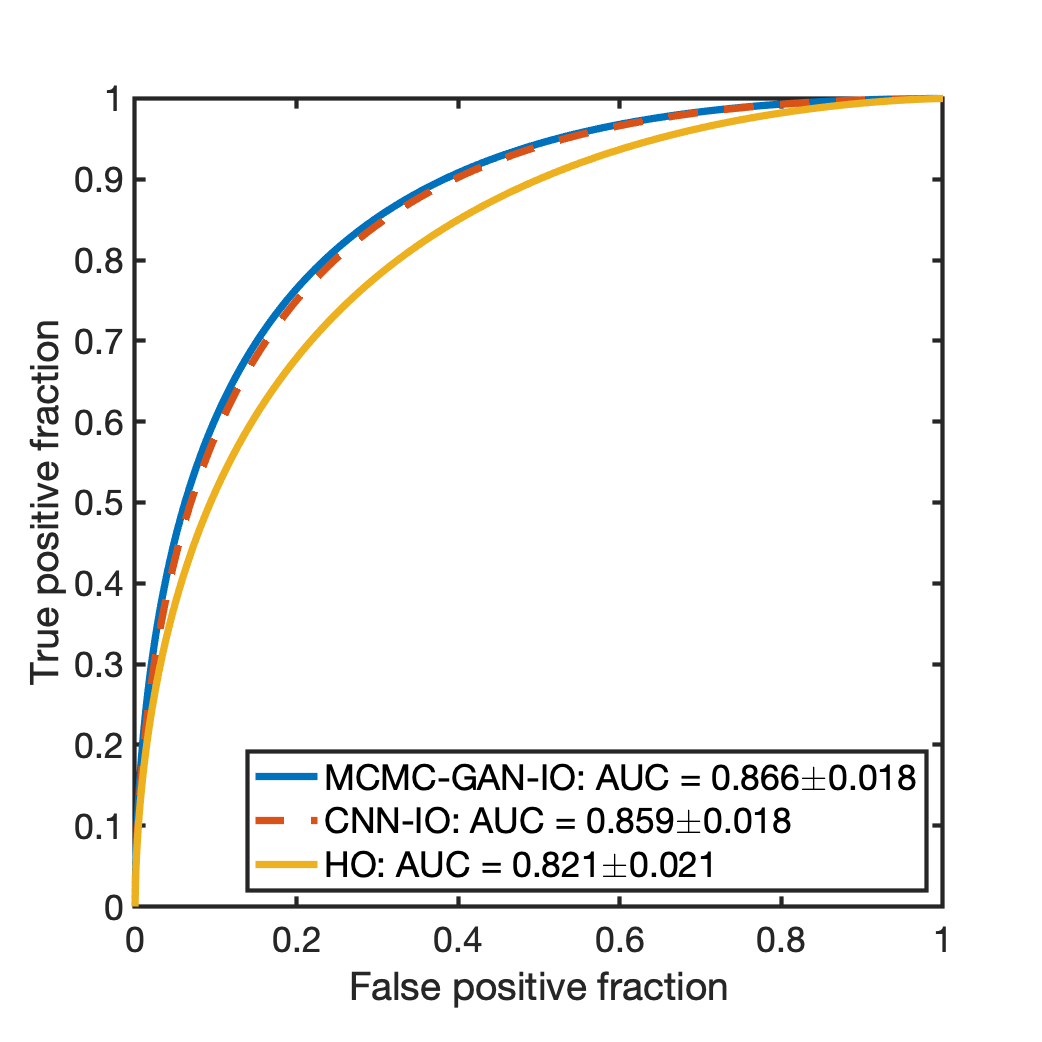}
 \caption{The ROC curves produced by the MCMC-GAN-IO, CNN-IO and the HO. The ROC curve corresponding to the MCMC-GAN IO is in close agreement with the CNN-IO and is higher than the HO.}
 \label{fig:MR_roc}
\end{figure}

The ROC curves corresponding to the MCMC-GAN IO (blue curve), CNN IO (red-dashed curve) and the HO (yellow curve) are shown in Fig. \ref{fig:MR_roc}.
The AUC value corresponding to the MCMC-GAN IO, CNN IO and the HO are $0.866\pm {0.018}$, $0.859\pm {0.018}$ and $0.821\pm {0.021}$, respectively.
The performance of the MCMC-GAN IO is in close agreement with that of the CNN IO and is higher than that of the HO.

 \section{Discussion and Conclusion}
 \label{sec:con}
The Bayesian Ideal Observer (IO) that employs complete task-specific information has been widely accepted for use in the evaluation and optimization of medical imaging systems.
 However, analytical computation of the IO has been limited to special cases that are rarely seen in applications of practical interest. 
 To address this need,  Kupinski \emph{et al.} proposed a sampling-based method that employs MCMC techniques to numerically compute the IO for lumpy object models  \cite{kupinski2003ideal}.
 This method has been also adapted to some other SOMs such as a binary texture model \cite{abbey2008ideal} and a parametrized torso phantom \cite{he2008toward}.
 However, the applicability of the MCMC method to more sophisticated SOMs that can represent object variability in realistic medical images remains under-explored.
In this work, we proposed a novel sampling-based method that employs MCMC techniques with  GAN-established-SOMs (MCMC-GAN) to approximate the IO.
Because the proposed MCMC-GAN method can be potentially employed with any SOMs established by GANs, the applicability of MCMC techniques to compute the IO is extended.
To demonstrate this, the MCMC-GAN  was applied to a set of clinical brain MR images in a numerical study that cannot be accomplished by the traditional MCMC method.
 
This study considered binary signal detection tasks in which the IO test statistic is described by the likelihood ratio.
The proposed MCMC-GAN method can also be applied to joint signal detection-localization tasks in which the IO is described by a modified generalized likelihood ratio test \cite{khurd2005decision}.
Moreover,
the MCMC-GAN method may also be employed to compute other quantities that can be described by Monte Carlo integration.
Another potential application of the MCMC-GAN method is to approximate the IO for joint signal detection-estimation tasks in which 
a quantity, known as the utility weighted posterior mean, needs to be computed by use of MCMC-based methods \cite{li2021supervised}.
It will be important to explore the ability of the MCMC-GAN to approximate the IO for joint 
signal detection-estimation tasks in the future.

There remain additional topics for future investigation. In this study, the samples of Markov chains were obtained by use of a pCN proposal \cite{cotter2013mcmc}. 
More advanced MCMC algorithms such as Metropolis adjusted Langevin algorithms (MALA) and Hamiltonian Monte Carlo (HMC) \cite{pereyra2015survey} 
can be readily implemented in our proposed MCMC-GAN framework.
This is possible because the gradient of the GAN-establish-SOM, which is a neural network, with respect to the latent vector can be readily computed on machine learning platforms such as Tensorflow~\cite{abadi2016tensorflow}.
It will be important to investigate the performance of the MCMC-GAN that employs other MCMC samplers. 
Moreover, one may also employ other advanced GAN methods such as StyleGANs \cite{karras2019style,karras2019analyzing,karras2021alias} and variational autoencoders \cite{kingma2013auto} to establish SOMs to be used in the MCMC-GAN method.

One limitation of the proposed MCMC-GAN method is that it 
requires the use of a GAN that can accurately sample from the true object distribution.
However,  currently there is no comprehensive way to assess GANs and the development of such assessment procedures within the context of medical imaging is a topic of ongoing research \cite{varun2023assessing}. As such, it remains unclear if GANs can reliably capture the object statistics that are required by the IO, which vary by task. However, even without a validated generative model, the proposed MCMC-GAN method can still be applied for signal detection tasks but it may only provide an approximation to the IO.

\bibliography{references}{}
\bibliographystyle{IEEETran}
\end{document}